\documentclass[
reprint,
superscriptaddress,
nofootinbib,
amsmath,amssymb,
aps,
prd,
showkeys,
nolongbibliography,
]{revtex4-2}

\usepackage{graphicx}
\usepackage{dcolumn}
\usepackage{bm}
\usepackage{multirow}
\usepackage{xcolor}
\usepackage[
    colorlinks=true,
    linkcolor=red,
    citecolor=blue,
    urlcolor=blue
]{hyperref}
\usepackage[inline]{enumitem}
\usepackage{booktabs}
\usepackage{soul}
\usepackage{amsmath,amssymb,amsfonts}
\usepackage{tikz}
\usepackage[percent]{overpic}

\newcommand{\LCDM}{$\Lambda$CDM}

\newcommand{\Om}{\Omega_m}
\newcommand{\Ol}{\Omega_{\ell}}

\definecolor{mygray}{RGB}{120,120,120}
\definecolor{dodgerblue}{RGB}{30,144,255}
\definecolor{tomato}{RGB}{255,99,71}

\newcommand{\draftfigure}[2][]{%
    \IfFileExists{#2}{\includegraphics[#1]{#2}}{%
    \fbox{\parbox[c][3.2cm][c]{0.92\linewidth}{\centering Missing figure:\\[-1mm]\texttt{\detokenize{#2}}}}}}

\begin{document}

\title{Metastable Dark Energy on the Phantom Brane}

\author{Purba Mukherjee}
\email{purba@kasi.re.kr}
\affiliation{Korea Astronomy and Space Science Institute, 776 Daedeok-daero, Yuseong-gu, Daejeon 34055, Republic of Korea}
\author{Arman Shafieloo}
\email{shafieloo@kasi.re.kr}
\affiliation{Korea Astronomy and Space Science Institute, 776 Daedeok-daero, Yuseong-gu, Daejeon 34055, Republic of Korea}
\affiliation{University of Science and Technology, Daejeon 34113, Korea}
\author{Varun Sahni}
\email{varun@iucaa.in}
\affiliation{Inter-University Centre for Astronomy and Astrophysics, Post Bag 4, Ganeshkhind, Pune 411007, India}
\author{Yuri Shtanov}
\email{shtanov@bitp.kyiv.ua}
\affiliation{Bogolyubov Institute for Theoretical Physics, Metrologichna St. 14-b, Kiev 03143, Ukraine}

\date{\today}

\begin{abstract}
Recent DESI observations hint at a preference for dynamical dark energy with an equation of state (EoS) that crosses the phantom divide at $w=-1$. In this work, we investigate \textit{metastable} dark energy (DE) on the minimal phantom brane. 
 Braneworld effects lead to $w < -1$ at early times, whereas metastable decay of DE results in $w > -1$ at late times. We consider three metastable decay prescriptions on the phantom brane: an exponentially decaying DE component (M1), decay of DE into non-baryonic dark matter (M2), and decay into dark radiation (M3). Using CMB observations, DESI DR2 BAO measurements, and the DES-Dovekie Type Ia supernova sample, we compare these models with \LCDM, the Chevallier--Polarski--Linder (CPL) parametrization, and the minimal phantom-brane model with a non-decaying $\Lambda$. We find that metastable models on the brane are preferred over $\Lambda$CDM, their decay rates being $\Gamma/H_0=0.362\pm0.086$, $0.539\pm0.125$, and $0.483\pm0.112$ for M1, M2, and M3, respectively. Best fits of metastable models improve over \LCDM\ by $\Delta\chi^2_{\rm MAP}=-17.14$, $-18.76$, and $-18.22$, while the corresponding $\Delta{\rm DIC}$ values are $-8.66$, $-9.63$, and $-6.56$. 
 Braneworld screening drives effective phantom behaviour at $z\gtrsim0.4$, while DE decay drives the EoS toward non-phantom values closer to the present epoch, the phantom divide $w=-1$ being crossed at $z \simeq 0.3$--$0.4$. 
 These results show that the interplay between metastable decay and phantom-brane screening provides a simple physical mechanism for an effective crossing of the phantom divide consistent with DESI observations.
\end{abstract}


\maketitle

\section{Introduction}
\label{sec:intro}

The origin of the observed late-time acceleration remains one of the central open problems in cosmology \cite{Peebles:2003pk, Padmanabhan:2005cw, DiValentino:2025otz, CosmoVerseNetwork:2025alb}. In the standard 6-parameter \LCDM\ model, the accelerated expansion is attributed to a cosmological constant with equation of state (EoS) $w=-1$ \cite{Zeldovich:1968ehl, Weinberg:1988cp}. Although this minimal description is remarkably successful, the microscopic origin and small observed value of the cosmological constant remain unexplained \cite{Sahni:1999gb, Padmanabhan:2002ji}. It is therefore important to investigate alternatives in which the effective DE sector evolves with cosmic time \cite{Chevallier:2000qy, Linder:2002et, Copeland:2006wr, Hazra:2013dsx}.

Recent distance measurements, in particular baryon acoustic oscillation (BAO) data from DESI \cite{DESI:2025zpo, DESI:2025zgx} combined with Type Ia supernovae (SNe Ia) \cite{Scolnic:2021amr, DES:2025sig, Rubin:2026qdt, Hoyt:2026fve, Camilleri:2026rer}, cosmic microwave background (CMB) \cite{Planck:2018vyg, AtacamaCosmologyTelescope:2025nti} and weak lensing (WL) \cite{DES:2026jmi} information, have renewed interest in dynamical DE and in phenomenologies in which the effective EoS crosses the phantom divide, $w=-1$ \cite{DESI:2025fii, Mukherjee:2025ytj, Berti:2025phi, Ormondroyd:2025exu, DES:2025sig}. Such a crossing is difficult to realize with a single minimally coupled canonical scalar field \cite{Ratra:1987rm, Vikman:2004dc, Caldwell:1999ew, Caldwell:2003vq}. It might instead indicate a more complicated dark sector, modified gravity, or an effective description involving more than one physical mechanism \cite{Braglia:2025gdo,Argudo-Panes:2026qkw,Khoury:2025txd, Giare:2024smz, Shah:2025ayl, Chakraborty:2025syu, Mukherjee:2025myk, Wolf:2025acj, Naidoo:2026umv, Mishra:2025goj, Guedezounme:2025wav, Li:2026hwq, Gomez-Valent:2026ept, Parker:1968mv, Landim:2016isc, Jiang:2026cqh, Delaunay:2026fse, Delaunay:2026jto, Li:2026xaz, Li:2026asg, Escamilla:2026eks, Ginat:2026fpo, Liu:2025bss, SanchezLopez:2025uzw, RoyChoudhury:2025iis, Guedezounme:2025wav, Gomez-Valent:2025mfl, Arora:2025msq, Adam:2025kve, Wolf:2025acj, Koutroulis:2026gjr, Gomez-Valent:2026ept, Abdalla:2026sis, Mishra:2026tzn, Ren:2026jyw, Chanda:2026dmx}.

One possibility is \emph{metastable dark energy} \cite{Shafieloo:2016bpk}, in which the DE density $\rho_{x}$ follows a radioactive-like decay law,
\begin{equation}
    \dot\rho_{x}=-\Gamma\rho_{x},
    \label{eq:radioactive-law}
\end{equation}
where $\Gamma$ is an intrinsic decay rate \cite{Li:2019san, Yang:2020zuk, Mukherjee:2026qzl}. For $\Gamma>0$, the DE density decreases with cosmic time, and its effective fluid description is quintessence-like. A second possibility is the \emph{phantom brane}, a branch of induced-gravity braneworld cosmology \cite{Shtanov:2000vr} in which the modified Friedmann equation screens the cosmological constant contribution, producing an effective $w_{\rm DE}<-1$ without introducing a pathological ghostlike scalar field with a wrong-sign kinetic term \cite{Deffayet:2000uy, Sahni:2002dx, Bag:2018jle, Bag:2021cqm}. 

This combination is particularly interesting because the two mechanisms can drive the effective DE dynamics in opposite directions. Braneworld screening tends to push the effective EoS below $-1$ at early times \cite{Mishra:2025goj}, whereas positive metastable decay pushes it above $-1$ at late times \cite{Mukherjee:2026qzl}. The resulting cosmology can therefore cross the phantom divide even though the metastable component itself remains non-phantom. This provides a concrete realization of a phantom-divide-crossing background without requiring the DE field responsible for the decay to have a phantom kinetic term.

In this work, we test this scenario using CMB \cite{Planck:2020olo, Rosenberg:2022sdy, Jense:2025wyg, AtacamaCosmologyTelescope:2025blo, SPT-3G:2025bzu} observations, DESI DR2 BAO \cite{DESI:2025fii} measurements, and the DES-Dovekie \cite{DES:2025sig} Type Ia supernova sample, and compare it with \LCDM, $w_0w_a$CDM \cite{Chevallier:2000qy, Linder:2002et}, and a minimal phantom-brane model with a cosmological constant \cite{Sahni:2002dx, Bag:2021cqm}. We consider three metastable DE decay prescriptions on the phantom brane: effective exponential decay, decay into dark matter (DM), and decay into dark radiation (DR). The decay rate is treated as an intrinsic property of the DE sector, rather than being fixed by the Hubble expansion rate or curvature. Beyond constraining the model parameters, we reconstruct the redshift evolution of the effective EoS. We assess the statistical performance of these models relative to standard \LCDM\ using $\Delta\chi^2_{\rm MAP}$ and the deviance information criterion (DIC) \cite{Liddle:2007fy, Grandis:2016fwl}. We find a preference for positive decay rates in all three metastable models, together with an improved fit relative to standard \LCDM. For each decay scenario, the metastable DE component admits a canonical scalar-field representation \cite{Mishra:2025goj}, while its interplay with phantom-brane screening produces an effective crossing of the phantom divide. Together, these results provide a viable physical explanation for the evolving DE behaviour indicated by DESI.

This paper is organized as follows. In Sec.~\ref{sec:brane}, we review the background dynamics of the phantom brane. In Sec.~\ref{sec:model}, we introduce the three metastable DE models. Section~\ref{sec:data} describes the observational data and analysis methodology. We present and discuss our results in Sec.~\ref{sec:results}. Finally, Sec.~\ref{sec:summary} summarizes our main conclusions.

\section{Braneworld Cosmology}
\label{sec:brane}

The brane, which is a 4D boundary of the 5D bulk space, is assumed to be spatially {\em marginally\/} closed (spatially flat in the limit), with no cosmological constant in the bulk (flat bulk space). It is filled with standard components (dark and baryonic matter and radiation with usual equation of state) and with dark energy, the model of which is not specified here. Our model is assumed to be described by the conventional action (in units $\hbar = c = 1$) \cite{Collins:2000yb, Dvali:2000hr, Shtanov:2000vr, Deffayet:2000uy, Sahni:2002dx}
\begin{equation}
S = \frac{M_p^3}{2} \left( \int_{\rm bulk} \!\! {\cal R} - 2 \int_{\rm brane} \!\! K \right) +  \frac{m_p^2}{2} \int_{\rm brane} \!\! R \ + \int_{\rm brane} \!\! L_\text{m} \, ,
\end{equation}
where ${\cal R}$ is the scalar curvature of the five-dimensional bulk, and $R$ is the scalar curvature corresponding to the induced metric $g_{\mu\nu}$ on the brane. The symbol $L_\text{m}$ denotes the Lagrangian density for all the four-dimensional physics whose dynamics is restricted to the brane (including dark matter and dark energy) so that it interacts only with the induced metric $g_{\mu\nu}$. The quantity $K$ is the trace of the symmetric tensor of extrinsic curvature of the brane. All integrations over the bulk and brane are taken with the corresponding natural volume elements. The universal constants $M_p$ and $m_p$ play the role of the five-dimensional and four-dimensional Planck masses, respectively. 

We normalise the brane Planck mass as $m_p^2 = 1 / 8 \pi G$, then the stress--energy tensor of matter is defined with the usual canonical factor of $1/2$, i.e., under variation of the metric,
\begin{equation}
\delta \int_{\rm brane} \!\!\! L_\text{m} = - \frac12 \int T_{\mu\nu} \delta g^{\mu\nu} \, .
\end{equation}
The fundamental crossover length scale in this theory is then defined as
\begin{equation} \label{ell}
\ell = \frac{2 m_p^2}{M_p^3} \, .
\end{equation}

The background cosmological evolution on the normal branch of the braneworld theory is described by \cite{Shtanov:2000vr, Deffayet:2000uy, Sahni:2002dx, Bag:2018jle}
\begin{align} \label{background-flat}
H^2 + \frac{\kappa}{a^2} &= \frac{\rho}{3 m_p^2} + \frac{2}{\ell^2}\left( 1 - \sqrt{1 + \frac{\ell^2 \rho}{3 m_p^2} } \right) \nonumber \\
&=  \left( \sqrt{\frac{1}{\ell^2} + \frac{\rho}{3 m_p^2}} - \frac{1}{\ell} \right)^2 \, .
\end{align}
We observe that, in the regime $H\gg \ell^{-1}$, our braneworld expands as in general relativity with the gravitational constant $8 \pi G = 1/m_p^2$, and with possible dark energy encoded in the total energy density $\rho$.

We consider a spatially flat Friedmann--Lema\^itre--Robertson--Walker geometry on the brane and define the normalized Hubble parameter as
\begin{equation}
    h(z)\equiv {H(z)}/{H_0},
\end{equation}
where $H_0$ denotes the present-day Hubble parameter. For the minimal phantom brane\footnote{A completely general expression for the expansion rate also contains contributions from a five-dimensional `bulk' cosmological constant and a dark radiation term -- corresponding to projections of the 5D Weyl tensor onto the brane. For simplicity both these terms have been set to zero in our treatment and the resulting braneworld is therefore referred to as the `minimal phantom brane'.} with a brane cosmological constant, the dimensionless expansion rate is
\begin{align}
\begin{split}
h^2(z)={}\Om(1+z)^3+\Omega_r(1+z)^4+\Omega_{\Lambda}+2\Ol ~~~~~~\\
~~~-2\sqrt{\Ol}\sqrt{\Om(1+z)^3+\Omega_r(1+z)^4+\Omega_{\Lambda}+\Ol} \, .
\end{split}
\label{eq:phantom-brane}
\end{align}
Here, the definition of the conventional Omega parameters is standard, while the braneworld scale is parametrized by
\begin{equation}
    \Ol=\frac{1}{\ell^2H_0^2}\, ,     \label{eq:omegal}
\end{equation}
where $\ell$ is the crossover length scale \eqref{ell} associated with the five-dimensional theory. The limit $\Ol\rightarrow0$ recovers the standard spatially flat Friedmann equation.

A key feature of the phantom branch is the {\em screening term} proportional to $-\sqrt{\Ol}$. This term reduces the value of $h(z)$ relative to $\Lambda$CDM at moderate values of $z$ 
and is responsible for the phantom-like properties of this braneworld. This can be easily seen by absorbing the non-GR terms in (\ref{eq:phantom-brane}) into an effective DE density,
\begin{equation}
\Omega_{\rm DE}(z)=h^2(z)-\Om(1+z)^3-\Omega_r(1+z)^4\, ,
 \label{eq:omega-de-eff}
\end{equation}
from which the effective EoS can be written as
\begin{equation}
 w_{\rm DE}(z)=-1+\frac{1+z}{3}\frac{{\rm d}\ln\Omega_{\rm DE}(z)}{{\rm d}z} \, .
 \label{eq:weff}
\end{equation}
For the phantom branch, the screening term produces an effective $w_{\rm DE}<-1$ without introducing a phantom matter component. The standard \LCDM\ expansion history is recovered in the limit $\Omega_\ell\rightarrow0$. 

Equation~\eqref{eq:phantom-brane} therefore defines the baseline braneworld model considered in this work. We now replace its constant vacuum contribution with a metastable DE component
, whose density 
evolves as a function of cosmic time, while retaining the same braneworld modification to the expansion rate.

\section{Metastable Models}
\label{sec:model}

\subsection*{M1: Exponentially decaying Dark Energy}
\label{subsec:model1}

Metastable DE density, $\rho_x$ in Eq.~\eqref{eq:radioactive-law}, obeys a radioactive decay law
\begin{equation}
\rho_{x}(t)=\rho_{x,0}\exp\left[-\Gamma(t-t_0)\right]\, ,
\label{eq:rho-time}
\end{equation}
where $\rho_{x,0}\equiv\rho_{x}(t_0)$ is its present value. 
Note that $\Gamma$ has dimensions of inverse-time and is the only free parameter in the model. The decay `half-life' is simply $t_{1/2} = \log{(2)}/\Gamma$ when $\Gamma > 0$. 
The lookback time is related to redshift through 
\begin{equation}
t_0-t=\int_0^z\frac{{\rm d}z'}{(1+z')H(z')}\, ,
\label{eq:lookback}
\end{equation}
which holds independent of the specific expansion history. Combining these relations and expressing the DE density in units of the present critical density, we obtain
\begin{equation}
\Omega_{x}(z)=\Omega_{{x},0}
\exp\left[\frac{\Gamma}{H_0}\int_0^z
\frac{{\rm d}z'}{(1+z')h(z')}\right]\, .
\label{eq:omegaDE}
\end{equation}
Thus, the dimensionless ratio $\Gamma/H_0$ sets the decay timescale relative to the present Hubble time. For $\Gamma>0$, the decay timescale is $\Gamma^{-1}$, while $\Gamma=0$ recovers a constant vacuum-energy density.

The expression \eqref{eq:omegaDE} is completely general and is therefore valid both in general relativity (GR) and in modified gravity. On replacing the constant contribution $\Omega_\Lambda$ in Eq.~\eqref{eq:phantom-brane} with the evolving DE density in Eq.~\eqref{eq:omegaDE} yields
\begin{align}
\begin{split}
h^2(z)={}&\Omega_{{x}}(z) +\Om(1+z)^3+\Omega_r(1+z)^4 \, \\ 
&+ 2\Ol -2\sqrt{\Ol}\left[\Ol+\Om(1+z)^3   \right. \\ 
&+ \left. \Omega_r(1+z)^4+  \Omega_{{x}}(z)\right]^{1/2} \, .\\
\end{split}
\label{eq:master}
\end{align}
The normalization condition $h(0)=1$ fixes the present-day metastable DE density parameter as
\begin{equation}
 \Omega_{{x},0}=1-\Om-\Omega_r+2\sqrt{\Ol} \, .
 \label{eq:closure}
\end{equation}
Equation~\eqref{eq:master} implicitly determines the expansion history, since $h(z)$ also enters the integral governing DE evolution. We therefore solve the coupled evolution of the DE density and the expansion rate self-consistently.
\noindent Two limiting cases provide useful consistency checks: 
\begin{enumerate}[left=0pt]
\item For $\Gamma=0$, Eq.~\eqref{eq:master} reduces to the minimal phantom-brane model with a cosmological constant.
\item For $\Omega_\ell=0$, $\Gamma\neq0$, we recover metastable DE in GR, \begin{equation}
h^2(z)=\Omega_{x}(z)+\Omega_{m}(1+z)^3 +\Omega_{r}(1+z)^4.
\label{eq:gr-metastable}
\end{equation}
In this limit, describing the metastable DE component as a conserved fluid gives
\begin{equation}
w_{x}(z)=-1+\frac{\Gamma}{3H(z)}.
\label{eq:wde-gr}
\end{equation}
Thus, for $\Gamma>0$ in an expanding universe, $w_{\rm DE}(z) \equiv w_{x}>-1$ and the metastable component alone cannot cross the phantom divide \cite{}. This conclusion remains valid for an intrinsic metastable component on the brane when its decay is absorbed into a conserved-fluid description. On the phantom brane, however, the total effective EoS, $w_{\rm DE}$ in Eq.~\eqref{eq:weff}, additionally includes gravitational screening. Its interplay with DE decay can therefore drive the effective EoS to cross the phantom divide, $w_{\rm DE}=-1$.
\end{enumerate}

\subsection*{M2: Decay into Dark Matter}
\label{subsec:model2}

We next consider the decay of metastable DE into non-baryonic dark matter (DM), governed by the coupled evolution equations
\begin{align}
\label{eq:decay-dm}
\begin{split}
\dot{\rho}_{x} &= -\Gamma\rho_{x}\, , \\
\dot{\rho}_{\rm DM} + 3H\rho_{\rm DM} &= \Gamma\rho_{x} \, .
\end{split}
\end{align}
Baryons and radiation remain separately conserved. The corresponding expansion rate on the phantom brane can be written as
\begin{align}
\begin{split}
h^2(z)&={}\Omega_{x}(z)+\Omega_{\rm DM}(z)
+\Omega_{b}(1+z)^3 \, \\
&+\Omega_{r}(1+z)^4+2\Omega_\ell -2\sqrt{\Omega_\ell} \left[\Omega_\ell+\Omega_{x}(z) \,\right.\\
&+\Omega_{\rm DM}(z)+\left.\Omega_{b}(1+z)^3+\Omega_{r}(1+z)^4\right]^{1/2}\, . \end{split}
\label{eq:brane-dm}
\end{align}
The normalization condition $h(0)=1$ gives
\begin{equation}
\Omega_{{x},0}
=1-\Omega_{{\rm DM},0}-\Omega_{b}-\Omega_{r} +2\sqrt{\Omega_\ell}.
\label{eq:closure-dr-2}
\end{equation}
For $\Gamma>0$, energy transfer from DE to DM modifies the usual $(1+z)^3$ scaling of the DM density and consequently the expansion history. The numerical implementation includes the associated modifications to the DM perturbation equations consistently, following Ref.~\cite{Mukherjee:2026qzl}.

\subsection*{M3: Decay into Dark Radiation}
\label{subsec:model3}

We also consider the decay of metastable DE into dark radiation (DR), governed by \begin{align}
\begin{split}
\dot{\rho}_{x}&=-\Gamma\rho_{x} \, ,\\
\dot{\rho}_{\rm DR}+4H\rho_{\rm DR}&=\Gamma\rho_{x}\, .
\end{split}
\label{eq:decay-dr}
\end{align}
Non-relativistic matter and radiation remain separately conserved. The dimensionless expansion rate on the phantom brane is then
\begin{align}
\begin{split}
&h^2(z)={}\Omega_{x}(z)+\Omega_{m}(1+z)^3
+\Omega_{\rm DR}(z)+2\Omega_\ell \\
&-2\sqrt{\Omega_\ell} \left[\Omega_\ell+\Omega_{x}(z)+\Omega_{m}(1+z)^3+\Omega_{\rm DR}(z)\right]^\frac{1}{2}\, . 
\end{split}
\label{eq:brane-dr}
\end{align}
Here, $\Omega_{\rm DR}(z)$ denotes the density of the decay-produced radiation in units of the present critical density, distinct from the standard radiation contribution. The DE density retains the form given in Eq.~\eqref{eq:omegaDE}, while the DR density is obtained by solving Eq.~\eqref{eq:decay-dr} together with the modified expansion equation \eqref{eq:brane-dr}. The normalization condition $h(0)=1$ gives
\begin{equation}
\Omega_{{x},0}
=1-\Omega_{{\rm DR},0}-\Omega_{m}-\Omega_{r} +2\sqrt{\Omega_\ell}.
\label{eq:closure-dr-3}
\end{equation}
For $\Gamma>0$, the DE decay continuously sources DR, whose density is diluted by the expansion. Both its background evolution and perturbation hierarchy are included consistently in the numerical implementation.

\section{Data and Analysis}
\label{sec:data}

We undertake Markov chain Monte Carlo (MCMC) inference with \textsc{Cobaya}\footnote{\url{https://github.com/CobayaSampler/cobaya}} \cite{Torrado:2020dgo}, using a modified \textsc{CLASS}\footnote{\url{https://github.com/lesgourg/class_public}} \cite{Blas:2011rf, Lesgourgues:2011re} that extends the metastable DE implementation of Ref.~\cite{Mukherjee:2026qzl} to include the phantom brane. We require MCMC chains to satisfy the Gelman-Rubin criterion $R-1<0.02$ and analyse the posteriors with \textsc{GetDist}\footnote{\url{https://github.com/cmbant/getdist}} \cite{Lewis:2019xzd}. We use the following data sets:
\begin{enumerate}[left=0pt]
\item \textbf{CMB:} Foreground-marginalised CMB temperature and polarization measurements from combined Planck, ACT, and SPT with CMB lensing data \cite{Jense:2025wyg}. We use Planck NPIPE CamSpec-lite{\footnote{\url{https://github.com/HTJense/camspec_npipe-lite}}} over $\ell=30$--$1500$ (TT), $30$--$1000$ (TE), and $30$--$600$ (EE), supplemented by the low-$\ell$ temperature \cite{Rosenberg:2022sdy} and SRoll2 \cite{Delouis:2019bub} polarization likelihoods{\footnote{\url{https://github.com/CobayaSampler/cobaya/tree/master/cobaya/likelihoods/planck_2018_lowl}}}. For ACT DR6, we use the CMB-only likelihood{\footnote{\url{https://github.com/ACTCollaboration/DR6-ACT-lite}}} over $\ell=1500$--$6500$ (TT), $1000$--$6500$ (TE), and $600$--$6500$ (EE) \cite{AtacamaCosmologyTelescope:2025blo}. We also include the SPT-3G D1 temperature and polarization likelihood through \textsc{candl}\footnote{\url{https://github.com/Lbalkenhol/candl}}$^{,}$\footnote{\url{https://github.com/SouthPoleTelescope/spt_candl_data}} \cite{Balkenhol:2024sbv} and the joint ACT DR6, Planck, and SPT-3G baseline lensing likelihood{\footnote{\url{https://github.com/qujia7/spt_act_likelihood}}} \cite{SPT-3G:2025bzu}.
\item \textbf{DESI:} The DESI DR2 BAO{\footnote{\url{https://github.com/CobayaSampler/bao_data/tree/master/desi_bao_dr2}}} \cite{DESI:2025wyn} consisting of transverse, radial, and volume-averaged distances relative to the sound horizon at the baryon drag epoch, $r_d$.
\item \textbf{DES:} The SN-Ia distance moduli from the DES-Dovekie 5-year sample{\footnote{\url{https://github.com/des-science/DES-SN5YR}}}, including the associated statistical and systematic uncertainties \cite{DES:2025sig}.
\end{enumerate}

We adopt CMB+DESI+DES as our primary data combination and compare it with CMB+DESI to assess the impact of the SN-Ia data. We vary the six cosmological parameters $\left\{H_0,\,\omega_b,\,\omega_{\rm cdm},\,\tau,\,\ln(10^{10}A_s),\,n_s\right\}$ common to all models, which define our baseline \LCDM\ analysis. We additionally vary $\left\{w_0,\, w_a\right\}$ for $w_0w_a$CDM, $\Omega_\ell$ for the minimal phantom-brane model with a cosmological constant, and $\left\{\Gamma/H_0,\,\Omega_\ell\right\}$ for the metastable models. For M2, $\omega_{\rm DM}$ replaces $\omega_{\rm cdm}$ and denotes the present-day physical dark-matter density, whose evolution includes the contribution from DE decay. For M3, we additionally report the derived parameter $\Omega_{{\rm DR},0}$, the present-day density parameter of dark radiation produced by DE decay. We assume spatial flatness, $\Omega_k=0$, and adopt $N_{\rm eff}=3.044$, with one massive neutrino species of mass $m_\nu=0.06,\mathrm{eV}$ and two effectively massless species.

\squeezetable
\begin{table*}[t]
\centering
\caption{Marginalized posterior constraints for CMB+DESI+DES. Two-sided uncertainties are quoted at $68\%$ CL and one-sided limits at $95\%$ CL.}
\label{tab:constraints}
\setlength{\tabcolsep}{5pt}
\renewcommand{\arraystretch}{1.2}
\resizebox{\textwidth}{!}{%
\begin{tabular}{lcccccc}
\toprule
\textbf{Parameter} & \textbf{$\Lambda$CDM} & \textbf{$w_0w_a$CDM} & \textbf{$\Lambda$CDM + brane} & \textbf{M1} & \textbf{M2} & \textbf{M3} \\
\midrule
$100\theta_\ast$ & $1.04177\pm 0.00020$ & $1.04168\pm 0.00020$ & $1.04168\pm 0.00020$ & $1.04165\pm 0.00020$ & $1.04157\pm 0.00020$ & $1.04164\pm 0.00019$ \\
$\omega_b$ & $0.022441\pm 0.000086$ & $0.022411\pm 0.000090$ & $0.022406^{+0.000093}_{-0.000084}$ & $0.022406\pm 0.000090$ & $0.022371\pm 0.000091$ & $0.022397\pm 0.000092$ \\
$\omega_{\rm cdm}$ & $0.11813\pm 0.00056$ & $0.11909\pm 0.00071$ & $0.11917^{+0.00067}_{-0.00075}$ & $0.11933\pm 0.00072$ & $-$ & $0.11961\pm 0.00076$ \\
$\omega_{\rm DM}$ & $-$ & $-$ & $-$ & $-$ & $0.178\pm 0.016$ & $-$ \\
$\tau$ & $0.0674\pm 0.0061$ & $0.0614^{+0.0053}_{-0.0060}$ & $0.0625^{+0.0056}_{-0.0064}$ & $0.0610\pm 0.0059$ & $0.0593^{+0.0053}_{-0.0060}$ & $0.0605^{+0.0052}_{-0.0061}$ \\
$\ln(10^{10}A_s)$ & $3.067\pm 0.011$ & $3.056\pm 0.011$ & $3.058^{+0.010}_{-0.012}$ & $3.055\pm 0.011$ & $3.052\pm 0.011$ & $3.055\pm 0.011$ \\
$n_s$ & $0.9739\pm 0.0035$ & $0.9720\pm 0.0037$ & $0.9722\pm 0.0036$ & $0.9719\pm 0.0036$ & $0.9703\pm 0.0037$ & $0.9713\pm 0.0037$ \\
\midrule
$\Gamma/H_0$ & -- & -- & -- & $0.362^{+0.094}_{-0.076}$ & $0.54\pm 0.13$ & $0.48\pm 0.11$ \\
$\Omega_\ell$ & -- & -- & $< 0.00326$ & $0.0103^{+0.0051}_{-0.0041}$ & $0.0140^{+0.0043}_{-0.0081}$ & $0.0130^{+0.0040}_{-0.0072}$ \\
$w_0$ & -- & $-0.806\pm 0.054$ & -- & -- & -- & -- \\
$w_a$ & -- & $-0.72^{+0.22}_{-0.20}$ & -- & -- & -- & -- \\
\midrule
$H_0\,\,[{\rm km\,s^{-1}\,Mpc^{-1}}]$ & $68.04\pm 0.23$ & $67.40\pm 0.56$ & $68.91\pm 0.43$ & $67.60\pm 0.54$ & $67.56\pm 0.55$ & $67.59\pm 0.54$ \\
$\Omega_{m}$ & $0.3050\pm 0.0031$ & $0.3130\pm 0.0054$ & $0.2995\pm 0.0038$ & $0.3116\pm 0.0051$ & $0.441\pm 0.038$ & $0.3123\pm 0.0051$ \\
$\Omega_{\rm DR}$ & $-$ & $-$ & $-$ & $-$ & $-$ & $0.092 \pm 0.024$ \\
$\sigma_8$ & $0.8160^{+0.0042}_{-0.0048}$ & $0.8160\pm 0.0072$ & $0.8229\pm 0.0052$ & $0.8099\pm 0.0060$ & $0.607^{+0.034}_{-0.045}$ & $0.8015\pm 0.0073$ \\
$S_8$ & $0.8228\pm 0.0062$ & $0.8334\pm 0.0068$ & $0.8222\pm 0.0061$ & $0.8254\pm 0.0062$ & $0.733^{+0.016}_{-0.020}$ & $0.8177\pm 0.0061$ \\
$r_d\,\,[{\rm Mpc}]$ & $147.53\pm 0.17$ & $147.30\pm 0.19$ & $147.29\pm 0.19$ & $147.25\pm 0.19$ & $147.18\pm 0.20$ & $147.18\pm 0.20$ \\
$h r_d\,\,[{\rm Mpc}]$ & $100.38\pm 0.41$ & $99.28\pm 0.83$ & $101.50\pm 0.63$ & $99.54\pm 0.80$ & $99.44\pm 0.81$ & $99.48\pm 0.78$ \\
\bottomrule
\end{tabular}}
\end{table*}

We obtain best-fit points with \textsc{iminuit}\footnote{\url{https://github.com/scikit-hep/iminuit}} \cite{iminuit, James:1975dr}, initialized at the maximum-a-posteriori (MAP) points from the MCMC chains. We assess the relative goodness of fit of each model with respect to \LCDM\ using the difference in likelihood deviance at their respective MAP points, 
\begin{equation}
 \Delta\chi^2_{\rm MAP}
 =\chi^2_{{\rm MAP},{\rm model}}
 -\chi^2_{{\rm MAP},\Lambda{\rm CDM}}.
 \label{eq:dchi2}
\end{equation}
To account for model complexity, we also compute the deviance information criterion (DIC) \cite{Liddle:2007fy, Grandis:2016fwl}, using
\begin{equation}
 {\rm DIC}=2\langle\chi^2\rangle-\chi^2_{\rm MAP},
 \label{eq:dic}
\end{equation}
where $\langle\chi^2\rangle$ denotes the posterior mean deviance, and define
\begin{equation}
 \Delta{\rm DIC} = {\rm DIC}_{\rm model}-{\rm DIC}_{\Lambda{\rm CDM}}.
 \label{eq:ddic}
\end{equation}
Note that negative $\Delta\chi^2_{\rm MAP}$ indicates an improved fit, while negative $\Delta{\rm DIC}$ indicates a preference relative to \LCDM\ under this criterion.

\begin{figure*}
    \centering
    \includegraphics[width=0.45\linewidth]{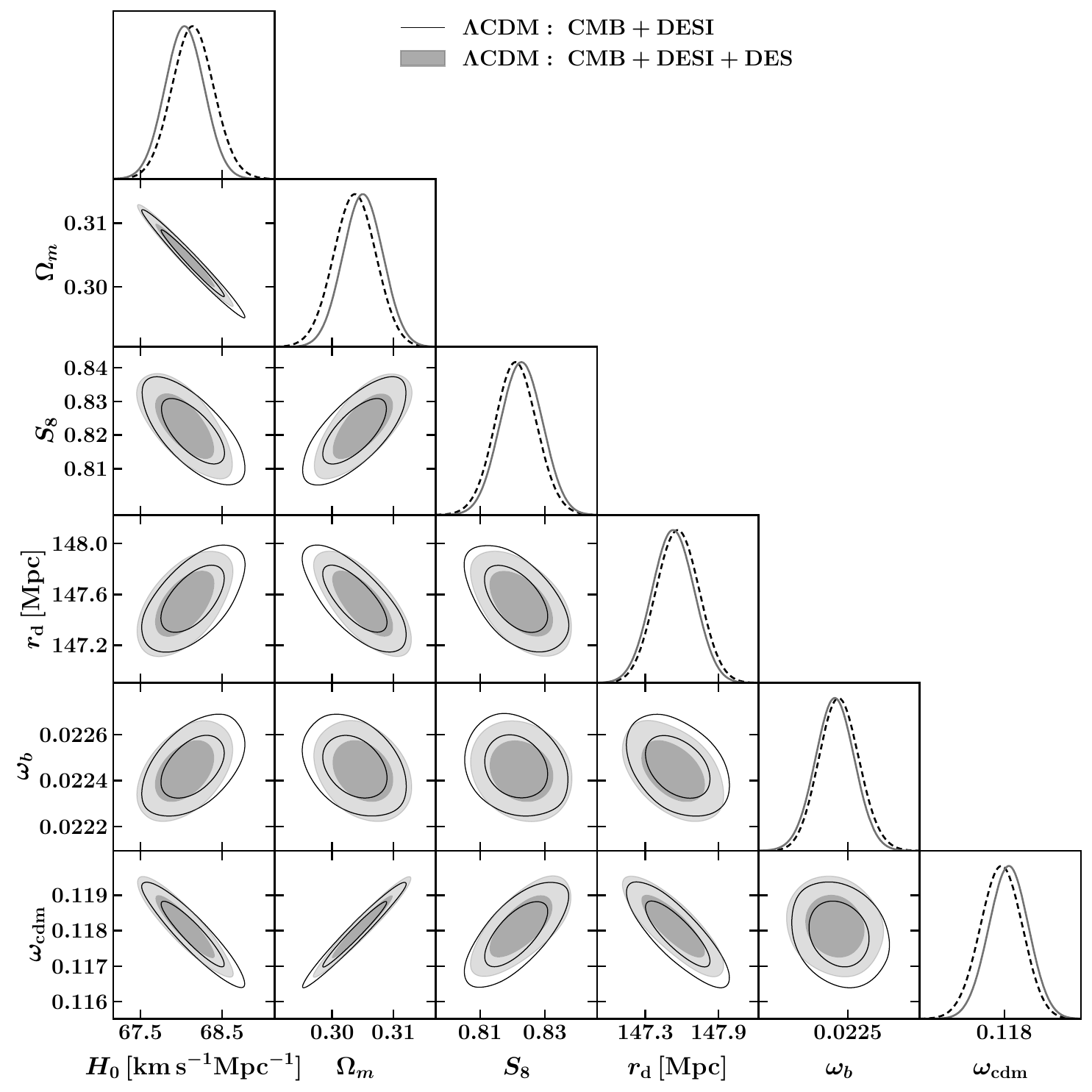}
    \includegraphics[width=0.45\linewidth]{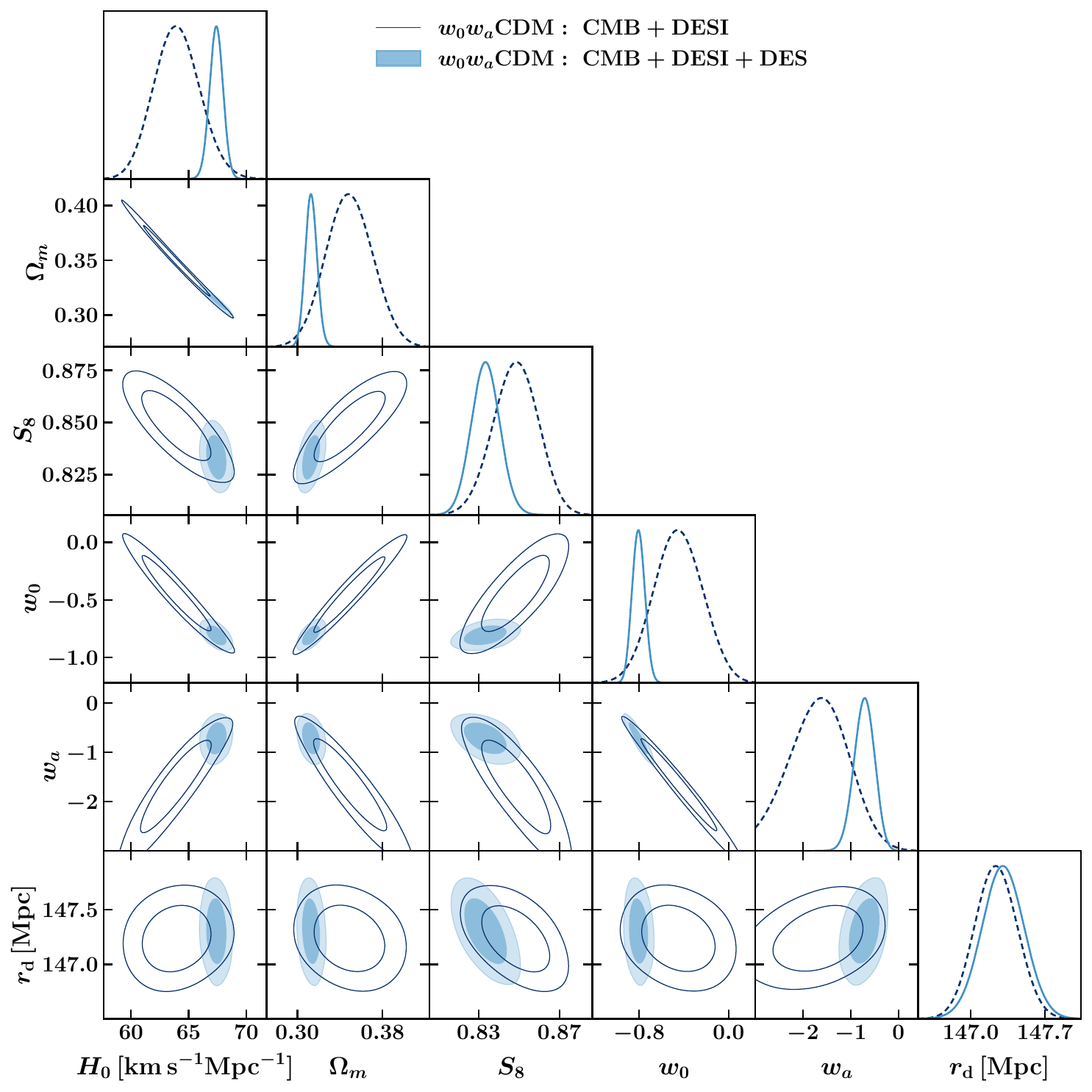}
    \includegraphics[width=0.45\linewidth]{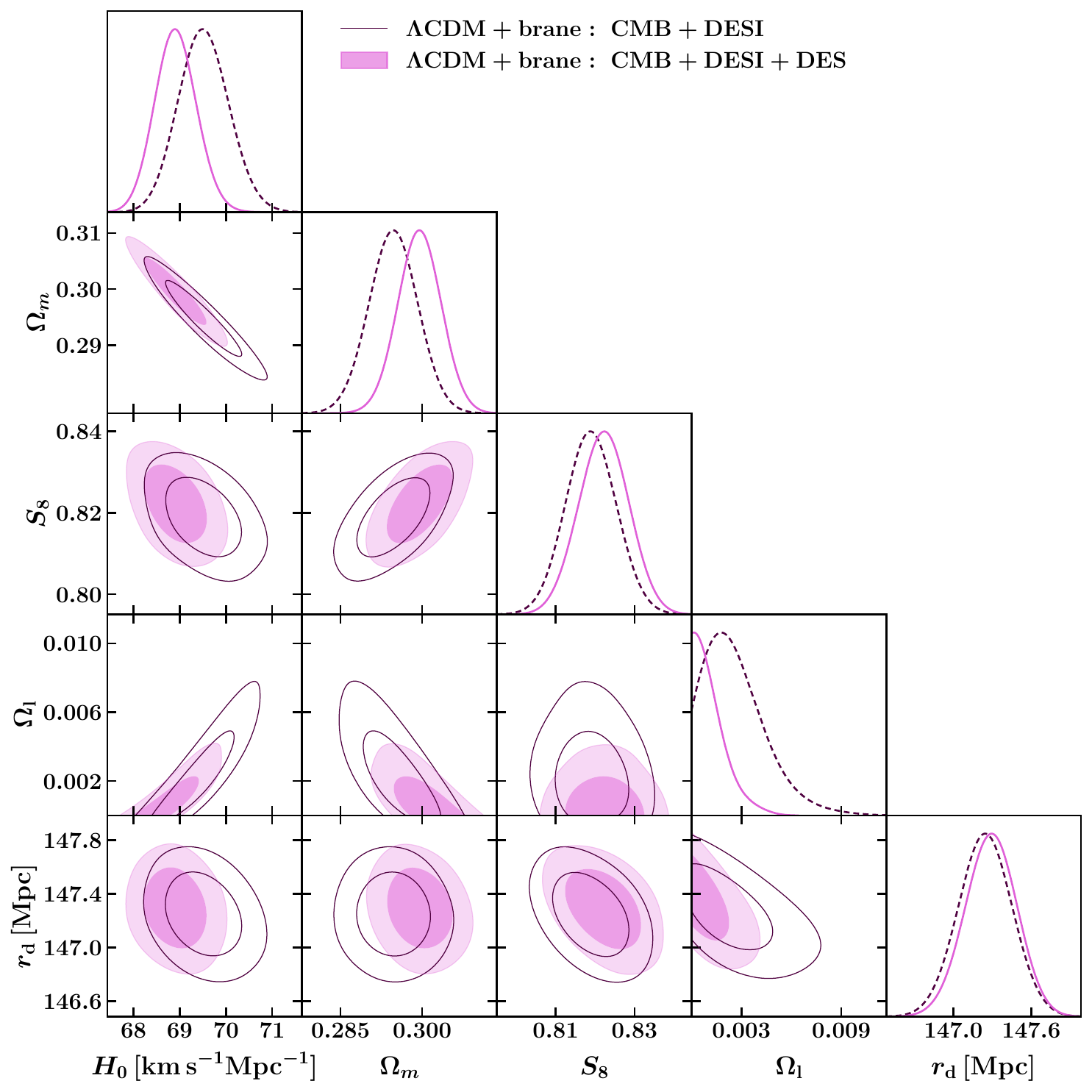}
    \includegraphics[width=0.45\linewidth]{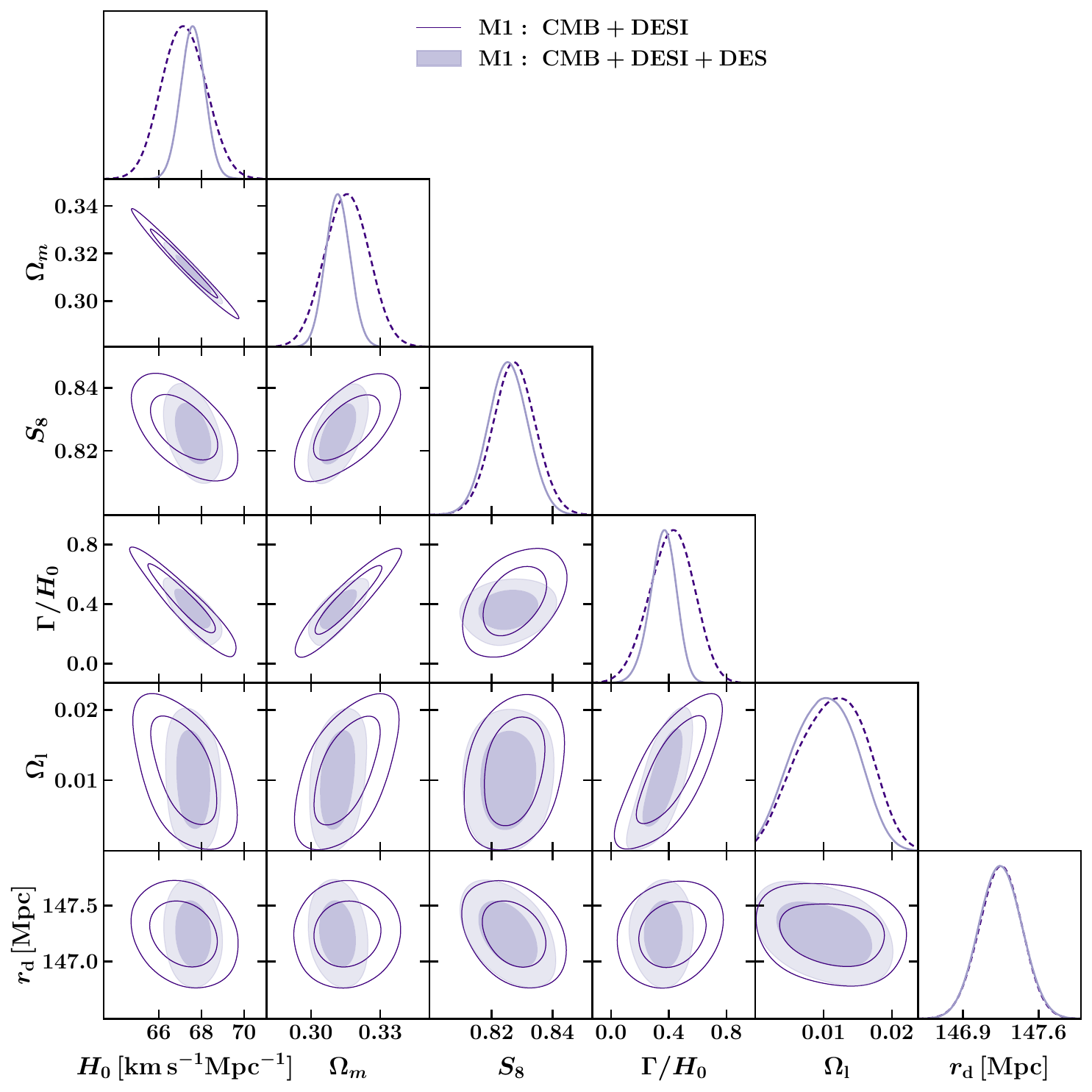}
    \includegraphics[width=0.45\linewidth]{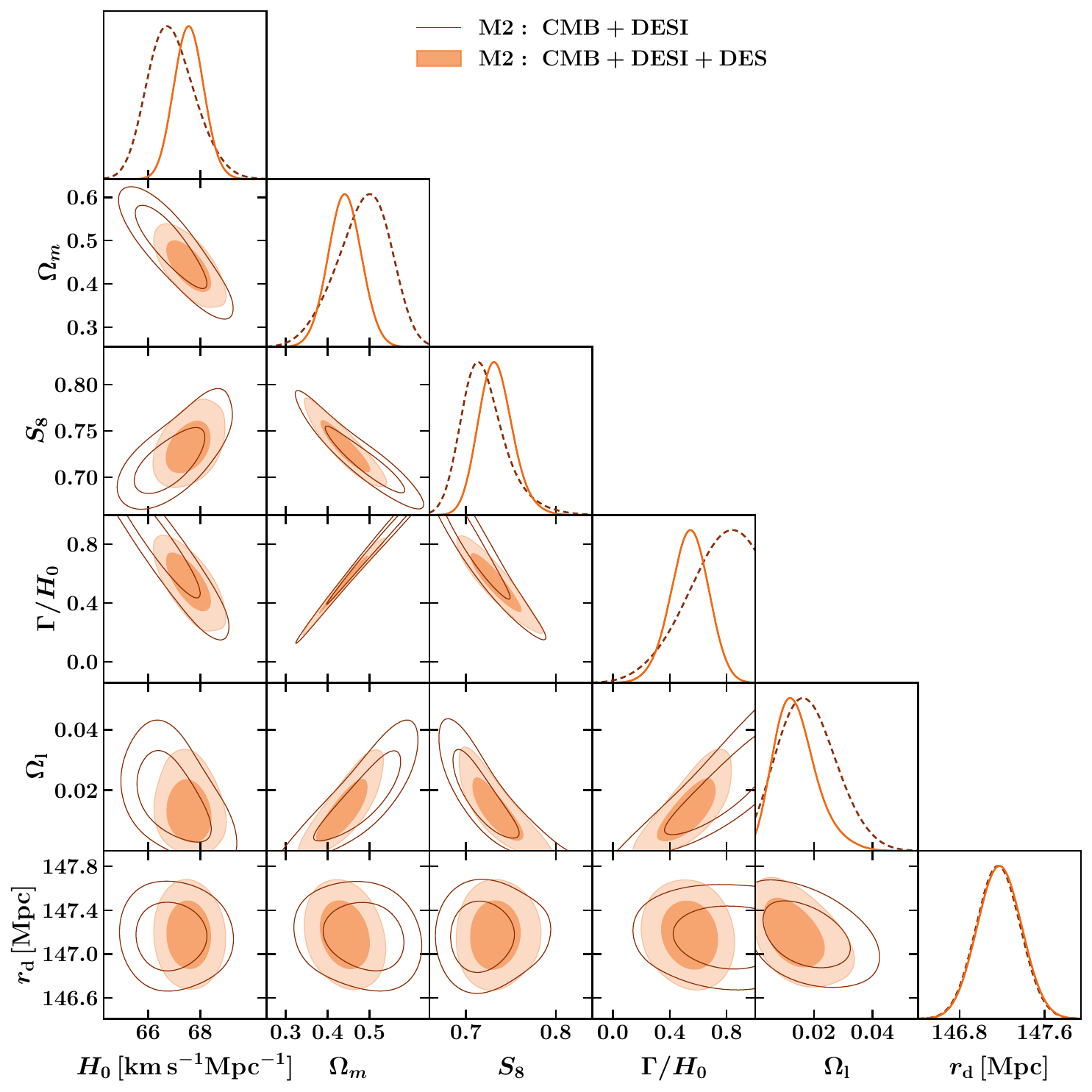}
    \includegraphics[width=0.45\linewidth]{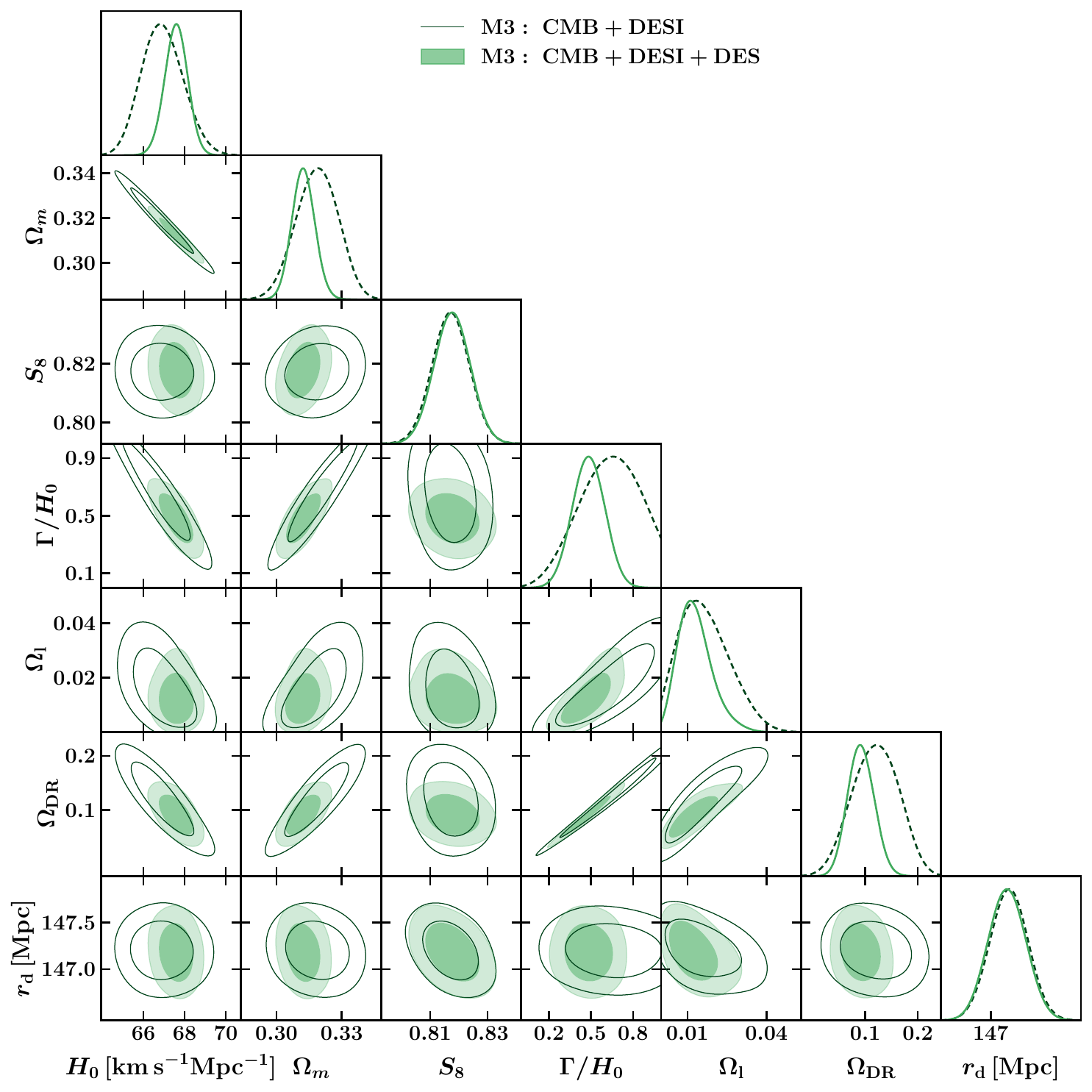}
    \caption{One- and two-dimensional marginalized posterior distributions for CMB+DESI and CMB+DESI+DES. The contours enclose the $68\%$ and $95\%$ credible regions. Note that $\Omega_\ell = 0$ corresponds to the GR limit of the phantom brane. $\Omega_\ell > 0$ is indicative of a preference for braneworld cosmology over $\Lambda$CDM.  }
\label{fig:triangle}
\end{figure*}

\section{Results and Discussion}
\label{sec:results}

We present the constraints on the three metastable DE models on phantom-brane and compare their statistical performance with \LCDM, $w_0w_a$CDM, and the minimal phantom-brane model with a cosmological constant. We quote two-sided marginalized constraints at the $68\%$ confidence level (CL) and one-sided limits at the $95\%$ CL. Table~\ref{tab:constraints} summarizes the marginalized parameter constraints, while Fig.~\ref{fig:triangle} shows the one- and two-dimensional posterior distributions obtained from the MCMC chains. We discuss the main results below.

\begin{enumerate}[left=0pt]
\item The combined CMB+DESI+DES data show a preference for positive decay rates in all three metastable models on the phantom brane. We find:
\begin{equation}
\frac{\Gamma}{H_0} \  = \ 
\begin{cases}
{0.362^{+0.094}_{-0.076}}\, \quad &\text{for \ M1}\, ,\\
{0.54\pm0.13}\, \quad &\text{for \ M2}\, ,\\
{0.48\pm0.11}\, \quad &\text{for \ M3}\, .
\end{cases}
\label{eq:gamma-results}
\end{equation}
These constraints indicate a metastable DE density that decreases with cosmic time. M1 gives a smaller central decay rate and a tighter constraint than the two daughter-production scenarios, although their marginalized posteriors overlap. Thus, on the phantom brane, observational data favor DE decay at approximately $4.8\sigma$, $4.2\sigma$, and $4.4\sigma$ CL for M1, M2, and M3, respectively.

\item The corresponding constraints on the brane parameter $\Omega_l$ are:
\begin{equation}
\Omega_\ell \ = \ 
\begin{cases}
0.0103^{+0.0051}_{-0.0041}\, \quad  &\text{for \ M1}\, ,\\
0.0140^{+0.0043}_{-0.0081}\, \quad  &\text{for \ M2}\, ,\\
0.0130^{+0.0040}_{-0.0072}\, \quad  &\text{for \ M3}\, ,
\end{cases}
\label{eq:brane-results}
\end{equation}
at $68\%$ CL. The quoted uncertainties suggest nonzero $\Omega_\ell$ at approximately $2.5\sigma$, $1.7\sigma$, and $1.8\sigma$ for M1, M2, and M3, respectively. By comparison, the minimal phantom-brane model with a cosmological constant gives $\Omega_\ell<0.00326$ at $95\%$ CL. Allowing DE decay therefore shifts the constraints toward larger $\Omega_\ell$, accommodating a stronger braneworld screening contribution to the expansion history.

\item Marginalized posterior constraints can be influenced by the volume of the remaining parameter space. We therefore also examine profile likelihoods, which isolate the best fit at each fixed parameter value and help assess whether the posterior preferences are supported by the likelihood itself. Figure~\ref{fig:profiles} shows the one-dimensional profiles obtained with \textsc{prospect}\footnote{\url{https://github.com/AarhusCosmology/prospect_public}} \cite{Holm:2023uwa} for $\Gamma/H_0$ and $\Omega_\ell$. We define
\begin{equation}
\Delta\chi^2_{\rm profile}(\theta) = \min_{ \boldsymbol{\eta}}\chi^2(\theta,\boldsymbol{\eta})
-\chi^2_{\min}\, ,
\label{eq:profile-definition}
\end{equation}
where $\theta$ denotes the profiled parameter and $\boldsymbol{\eta}$ represents the remaining parameters. Each profile is normalized to its corresponding model minimum, whose location is marked on the curve.

\begin{figure*}[t]
    \centering
    \includegraphics[width=0.35\linewidth]{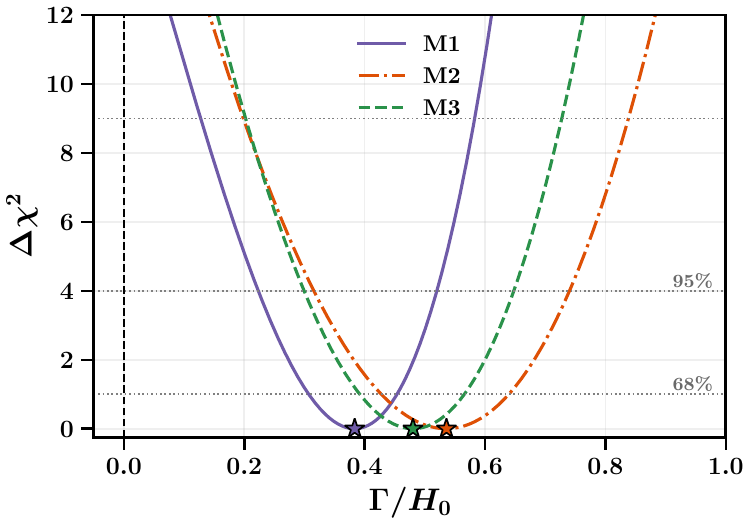}
    \includegraphics[width=0.35\linewidth]{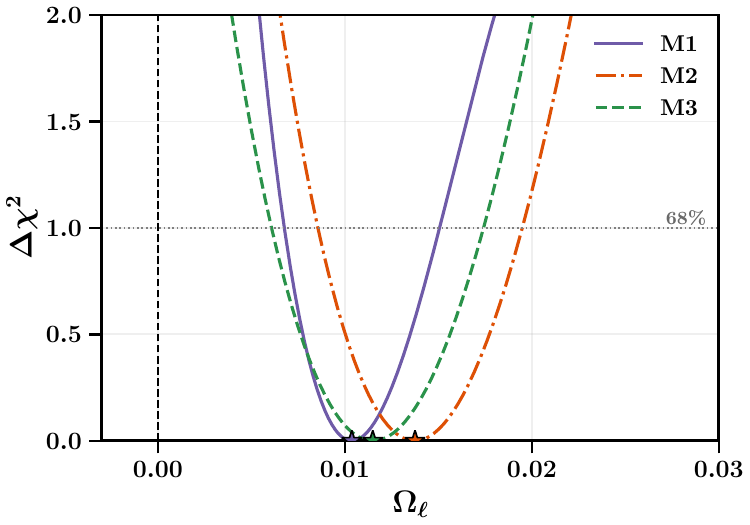}
    \caption{One-dimensional profile likelihoods for $\Gamma/H_0$ (left) and $\Omega_\ell$ (right), obtained with \textsc{PROSPECT} for M1, M2, and M3 using CMB+DESI+DES. Each profile is normalized to its corresponding model minimum, whose location is marked. The horizontal dotted lines indicate the approximate 68\% and 95\% CL shown in each panel. Note that $\Omega_\ell = 0$ corresponds to the GR limit of the phantom brane. $\Omega_\ell > 0$ is indicative of a preference for braneworld cosmology over $\Lambda$CDM.}
    \label{fig:profiles}
\end{figure*}

\begin{figure*}[t]
\centering
\includegraphics[width=\textwidth]{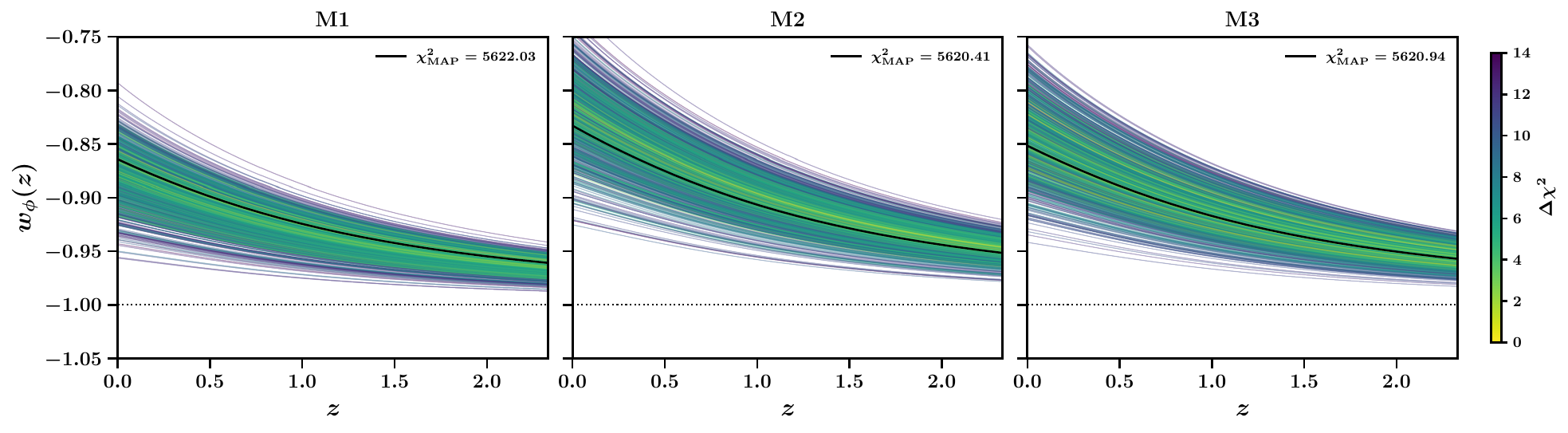}
\caption{Posterior histories of the metastable-component EoS, denoted $w_\phi\equiv w_x$ in the equivalent canonical scalar-field description, for M1--M3 in GR. Curves satisfy $\Delta\chi^2\leq14$, and the black line denotes the MAP history. Positive $\Gamma/H_0$ keeps the metastable component on the non-phantom side, $w_\phi>-1$. Note that while the metastable component on its own is quintessence-like with $w_\phi > -1$,  the presence of the phantom-brane reduces the effective EOS allowing it to cross the phantom divide $w_{\rm DE}=-1$ at $z \sim 0.5$, as shown in figure \ref{fig:weff}.}
\label{fig:wphi}
\end{figure*}

The decay-rate profiles have minima at positive $\Gamma/H_0$, consistent with the preference seen in the marginalized posteriors. M1 has the narrowest profile, while M2 and M3 permit a broader range of decay rates. The $\Omega_\ell$ profiles also have minima at positive values of order $10^{-2}$, with M2 preferring a somewhat larger value than M1 and M3. These results support the posterior preference for both DE decay and braneworld screening. Since each profile is normalized separately, the plotted minima do not compare the goodness of fit between models.

\item The inferred Hubble constant is $H_0\simeq67.6\,{\rm km\,s^{-1}Mpc^{-1}}$ for all three metastable models, close to the \LCDM\ value $68.04\pm0.23\,{\rm km\,s^{-1}Mpc^{-1}}$ and consistent with $67.40\pm0.56\,{\rm km\,s^{-1}Mpc^{-1}}$ for $w_0w_a$CDM. The minimal phantom-brane model with a cosmological constant instead gives a higher value of $68.91\pm0.43\,{\rm km\,s^{-1}Mpc^{-1}}$. The sound horizon remains nearly unchanged across the models, indicating a stable early-time acoustic calibration: $r_d$ varies only from about $147.18$ to $147.53\,\mathrm{Mpc}$, while differences in $h r_d$ mainly reflect shifts in $H_0$.

\begin{figure*}[t]
\centering
\includegraphics[width=\textwidth]{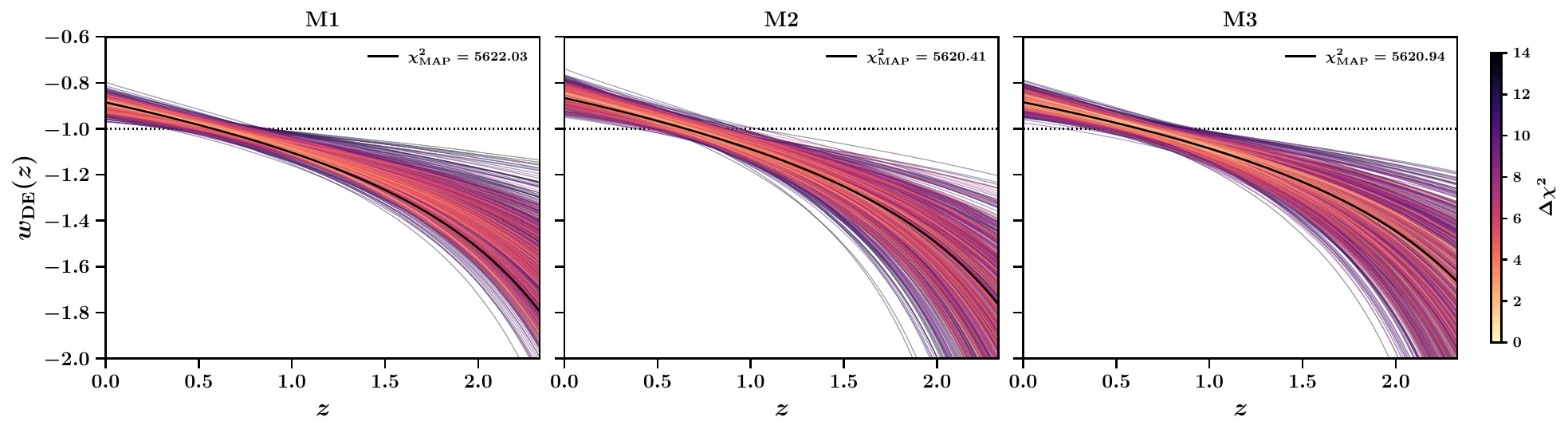}
\caption{Posterior histories of the total effective DE EoS $w_{\rm DE}(z)$ for M1--M3. Although the metastable component itself is non-phantom with $w_\phi > -1$, DE decay together with braneworld screening produce a crossing of the phantom divide $w_{\rm DE}=-1$ at $z \sim 0.5$. Curves satisfy $\Delta\chi^2\leq14$, and the black solid line denotes the MAP estimates.
}
\label{fig:weff}
\end{figure*}

\begin{figure*}[t]
\centering
\includegraphics[width=0.4\textwidth]{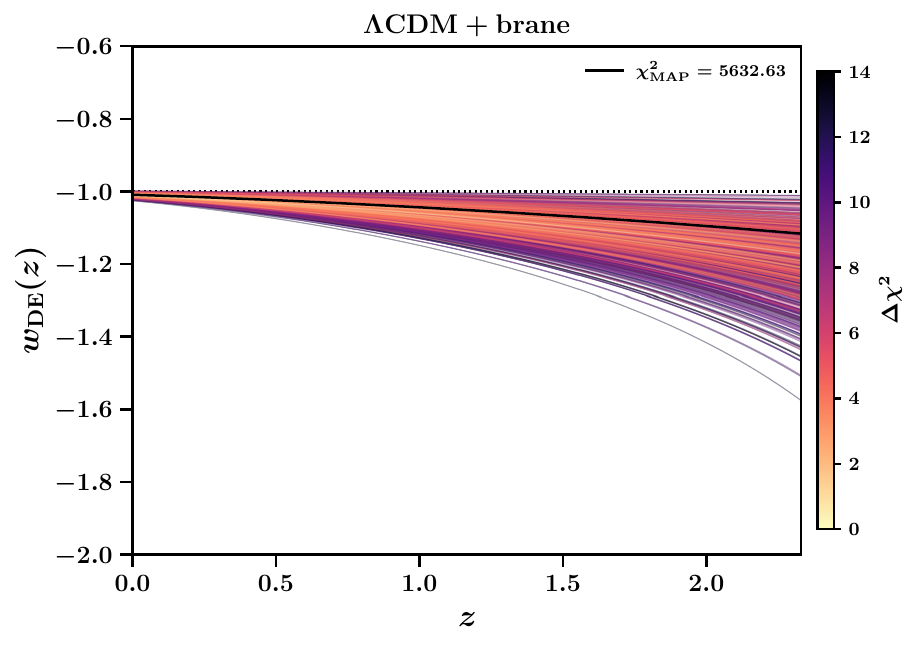}
\includegraphics[width=0.4\textwidth]{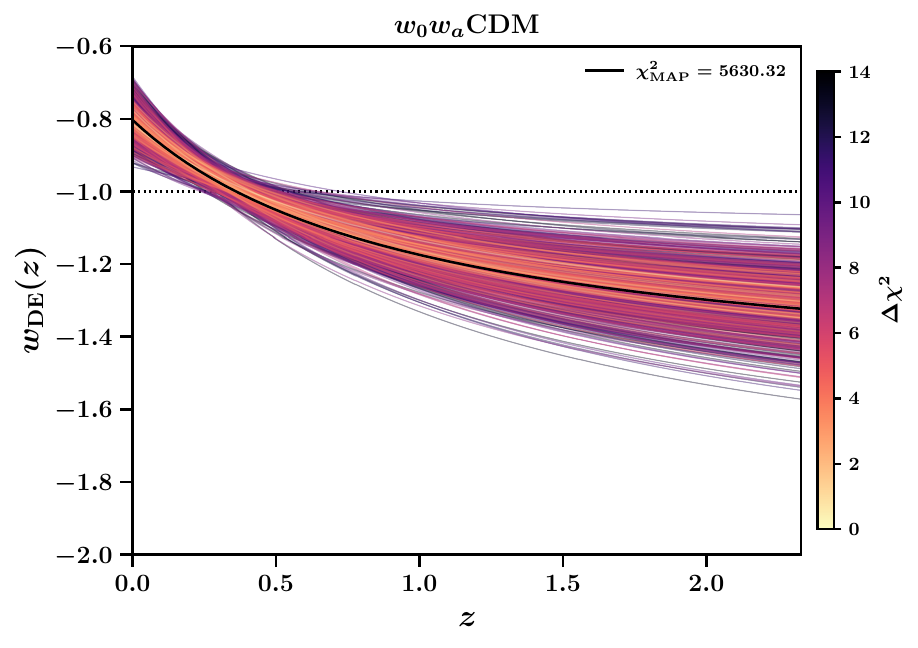}
\caption{
Posterior histories of $w_{\rm DE}(z)$ for the minimal phantom-brane model with a cosmological constant (left) and CPL (right), shown for comparison. Here $w_{\rm DE}$ denotes the total effective EoS for the brane model and the parametrized EoS for CPL.}
\label{fig:weff_cpl_brane}
\end{figure*}

\begin{figure*}[t]
\centering
\includegraphics[width=\textwidth]{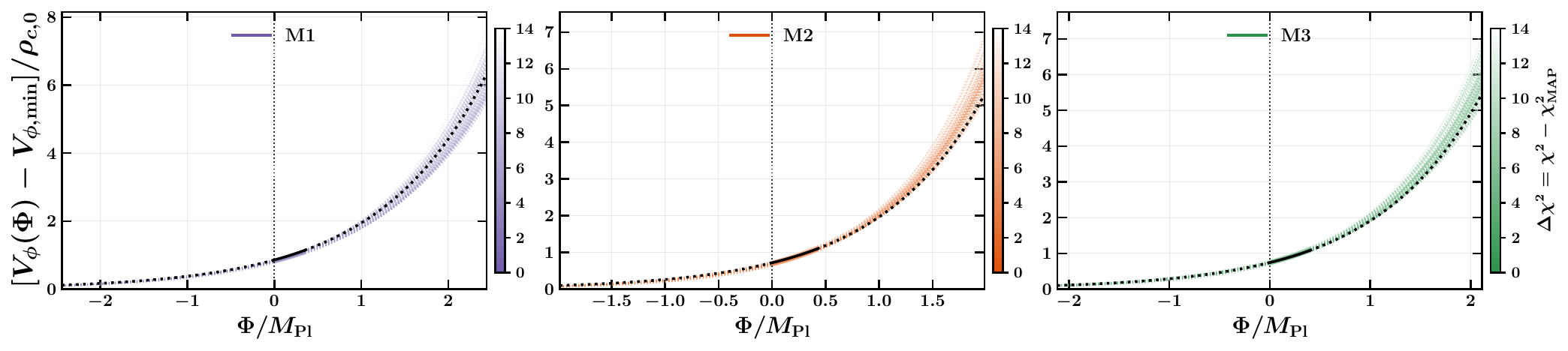}
\includegraphics[width=\textwidth]{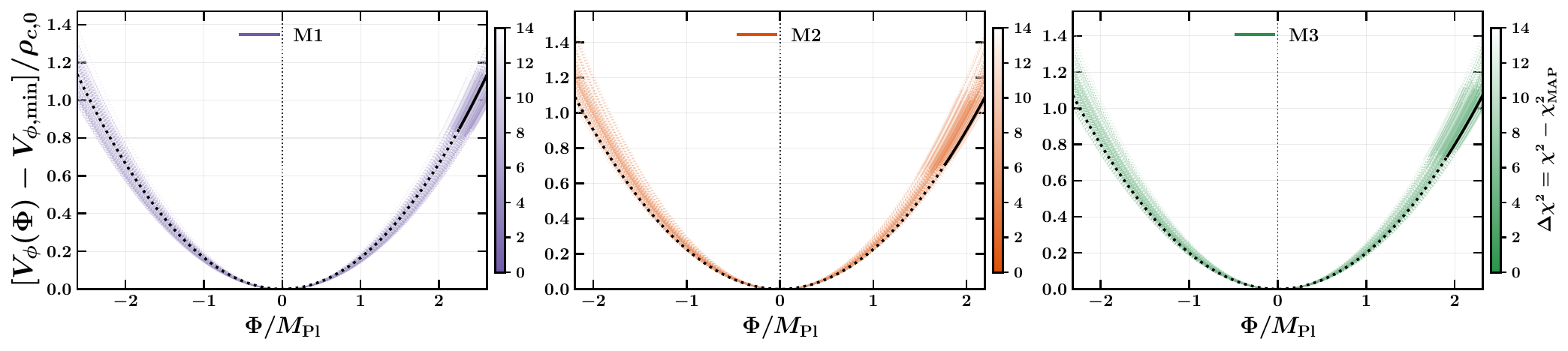}
\includegraphics[width=\textwidth]{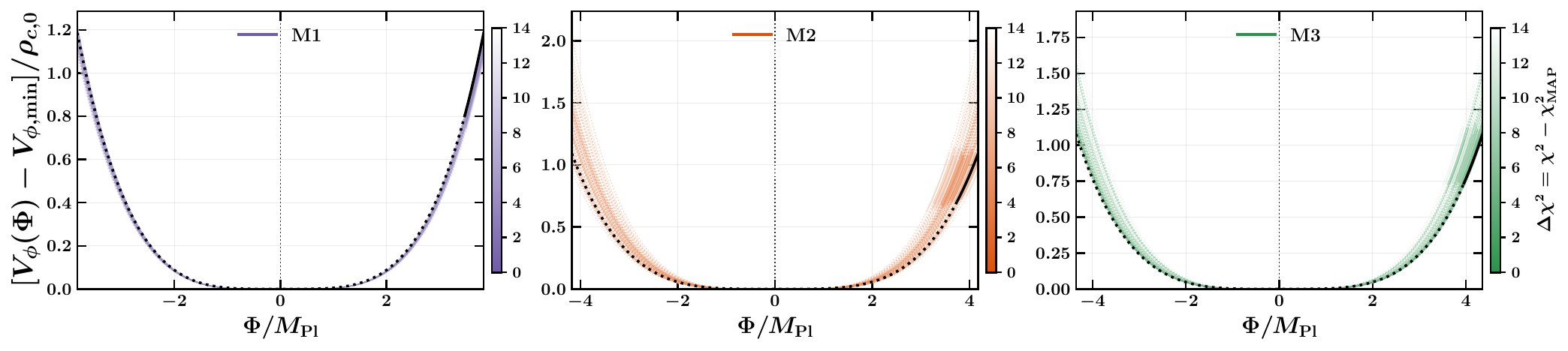}
\caption{Exponential (top), quadratic (centre), and quadratic (bottom) fits to the scalar-potential branches for M1--M3. Solid segments cover the reconstructed field intervals, while dashed segments show extrapolations. Colours encode $\Delta\chi^2$ relative to the MAP, and black curves show fits to the best-fits. For the exponential form, $V_{\phi,\min}=0$ denotes its asymptotic infimum.}
\label{fig:potential}
\end{figure*}

\item The matter and clustering constraints distinguish the decay channels more clearly. M1 and M3 give $\Omega_{m}=0.3116\pm0.0051$ and $0.3123\pm0.0051$, respectively. The DE-to-DM scenario instead gives $\Omega_{m}=0.441\pm0.038$, substantially higher than in the other models. For M1, M2, and M3, we obtain $\sigma_8=0.8099\pm0.0060$, $0.607^{+0.034}_{-0.045}$, and $0.8015\pm0.0073$, respectively, with corresponding values $S_8=0.8254\pm0.0062$, $0.733^{+0.016}_{-0.020}$, and $0.8177\pm0.0061$, all at $68\%$ CL. The larger present-day matter abundance is consistent with the transfer of energy from DE to DM. Together with the lower clustering amplitude, this distinguishes M2 from the other metastable models despite their similar inferred values of $H_0$. This distinct response is expected for an interacting DM-DE scenario and provides a direct way to separate M2 from the background-dominated M1 and the DR channel M3.

\item The DE $\longrightarrow$ DM decay channel in GR also shifts the matter density upward, giving $\Omega_{m}=0.341\pm0.023$ for \texttt{P-ACT+DESI+DES-Dovekie} in Ref.~\cite{Mukherjee:2026qzl}. Our phantom-brane analysis accommodates an even larger matter abundance, although the CMB combinations differ between the two studies. Moreover, Mukherjee \textit{et al}\cite{Mukherjee:2026qzl} showed that DESI DR1 full-shape \cite{DESI:2024mwx, DESI:2024jxi} measurements are particularly sensitive to DE decay into DM through its impact on the amplitude and scale dependence of galaxy clustering.  
The high matter density $\Omega_{m}$ and low clustering amplitude $S_8$ inferred for M2 motivate extending the phantom-brane analysis to full-shape galaxy clustering data. These measurements would test whether this parameter region is compatible with the observed growth of structure and whether the preference for DE decay to DM persists once we include both interaction and braneworld effects on clustering.

\item The reconstructed EoS histories distinguish the metastable component from the total effective DE sector. Describing the metastable component with decay rate $\Gamma$ through its equivalent conserved-fluid EoS gives
\begin{equation}
w_x(z)=-1+\frac{\Gamma}{3H(z)}.
\end{equation}
For positive density and $w_x>-1$, this description admits a canonical scalar-field representation. 
Hereafter, we denote $w_\phi\equiv w_x$, including in Fig.~\ref{fig:wphi}, to reflect the equivalent canonical scalar-field description. For a constant positive decay rate, $w_\phi(z)$ remains above $-1$ and approaches the vacuum value at higher redshift as $\Gamma/H(z)$ decreases. For interacting decay channels, this conserved-fluid description differs from the physical vacuum EoS of the decaying component.

\item The total effective EoS, $w_{\rm DE}$, additionally includes braneworld screening. As shown in Fig.~\ref{fig:weff}, the representative histories cross the phantom divide line $w_{\rm DE}=-1$ at $z\simeq0.3$--$0.4$: screening drives effective phantom behaviour at intermediate redshift, while DE decay produces non-phantom behaviour toward the present epoch. Thus, the combined sector can cross the phantom divide even though the conserved-fluid EoS assigned to the metastable component does not.

\item Figure~\ref{fig:weff_cpl_brane} shows the minimal phantom-brane model with a cosmological constant and CPL for comparison. We use $w_{\rm DE}$ for the total effective EoS of the braneworld models and for the parametrized CPL EoS, reserving $w_\phi$ for the metastable component's scalar-field representation. For comparison, CPL gives $w_0= -0.806\pm0.054$ and $w_a=-0.72^{+0.22}_{-0.20}$. The best-fit values describe a transition from phantom behaviour at earlier times to $w_{\rm DE}>-1$ today, consistent with the qualitative trend suggested by DESI and DES analyses \cite{DESI:2025zgx, DES:2026jmi, DES:2025sig}. Thus, the two scenarios, namely CPL and metastable DE on the phantom brane, exhibit similar phantom-divide crossings at the background level, but their origins differ: CPL prescribes the EoS evolution phenomenologically, whereas the metastable-brane models generate it through the competition between DE decay and braneworld screening.

\item The minimal phantom-brane model with a cosmological constant, also shown in Fig.~\ref{fig:weff_cpl_brane}, produces effective phantom behaviour through gravitational screening alone. Without DE decay, it does not exhibit the transition to $w_{\rm DE}>-1$ toward the present epoch found in the metastable-brane models.

\item We construct a canonical scalar-field representation of the metastable component for M1--M3 wherever $\rho_\phi\equiv\rho_x>0$ and $w_\phi\geq-1$. Writing
\begin{equation}
\rho_\phi=\frac{1}{2}\dot{\phi}^{\,2}+V_\phi,
\qquad
p_\phi=\frac{1}{2}\dot{\phi}^{\,2}-V_\phi,
\label{eq:scalar-rhop}
\end{equation}
with $p_\phi=w_\phi\rho_\phi$, we obtain
\begin{align}
\begin{split}
\frac{V_\phi(z)}{\rho_{c,0}}
&=\frac{1-w_\phi(z)}{2}\Omega_\phi(z)\, , \\
\left|\frac{{\rm d}(\phi/M_{\rm Pl})}{{\rm d}z}\right|
&=\frac{\sqrt{3[1+w_\phi(z)]\Omega_\phi(z)}}{(1+z)h(z)}\, .
\end{split}
\label{eq:scalar-map}
\end{align}
Here $\Omega_\phi(z)\equiv\rho_\phi(z)/\rho_{c,0}$,
$\rho_{c,0}=3M_{\rm Pl}^2H_0^2$, and
$M_{\rm Pl}=(8\pi G)^{-1/2}$. This representation reproduces the background density and pressure of the equivalent conserved fluid. For the metastable decay channels, it does not imply equivalence with the perturbations of an uncoupled canonical scalar field. 

We reconstruct $V_\phi(z)$ and $\phi(z)$ from the posterior samples and obtain $V_\phi(\Phi)$ parametrically over the field interval traced by each cosmological trajectory. Motivated by the scalar-field potentials studied on the phantom brane in Ref.~\cite{Mishra:2025goj}, we fit quadratic, quartic, exponential, and axion\footnote{The axion form corresponds to a pseudo-Nambu--Goldstone boson potential~\cite{Frieman:1995pm}.} forms to the same reconstructed branches:
\begin{align}
\begin{split}
V_{\rm quad}(\Phi)&=\frac{1}{2}m^2\Phi^2\, , \\
V_{\rm quart}(\Phi)&=\frac{\lambda_4}{4}\Phi^4 \, ,\\
V_{\rm expo}(\Phi)&=V_0\exp\left(\frac{\lambda\Phi}{M_{\rm Pl}}\right) \, , \\
V_{\rm axion}(\Phi)&=V_0\left[1-\cos\left(\frac{\Phi}{F}\right)\right]\, .
\end{split}
\label{eq:potential-forms}
\end{align}
Here, $\Phi$ denotes the field coordinate after a horizontal shift, which allows the reconstructed interval to occupy different portions of each candidate potential. We find that the exponential gives the closest match in every case. For M1, the remaining forms rank as quadratic, axion, and quartic; for M2 and M3, they rank as quartic, quadratic, and axion. The quadratic and axion forms perform almost equally well.

Figure~\ref{fig:potential} shows the exponential (top), quadratic (centre), and quartic (bottom) fits. These panels display the best-performing form in all three models alongside the second-ranked form for each decay channel. The quadratic form also describes the leading small-field behaviour of an axion potential,
\begin{equation}
V_{\rm axion}(\Phi)\simeq\frac{1}{2}m^2\Phi^2,
\qquad
m^2=\frac{V_0}{F^2},
\label{eq:axion-quadratic}
\end{equation}
for $|\Phi|\ll F$, as illustrated in Fig.~1 of Ref.~\cite{Mishra:2025goj}. Solid segments cover the reconstructed intervals, while dotted segments extend the fitted functions beyond them. The displayed minima and the behaviour outside the reconstructed intervals follow from the chosen functional forms, rather than independent constraints from the chains. The colour gradients retain the original metastable-chain $\Delta\chi^2$ values.

\begin{table}[t]
\centering
\caption{Model comparison relative to \LCDM\ for CMB+DESI+DES-Dovekie.}
\label{tab:modelcomparison}
\begin{tabular}{lrr}
\toprule
Model & $\Delta\chi^2_{\rm MAP}$ & $\Delta{\rm DIC}$ \\
\midrule
CPL & $-8.84$ & $-17.06$ \\
$\Lambda{\rm CDM}+{\rm brane}$ & $-6.54$ & $+5.14$ \\
M1 & $-17.14$ & $-8.66$ \\
M2 & $-18.76$ & $-9.63$ \\
M3 & $-18.22$ & $-6.56$ \\
\bottomrule
\end{tabular}
\end{table}

\begin{figure}[t]
\centering
\includegraphics[width=0.8\linewidth]{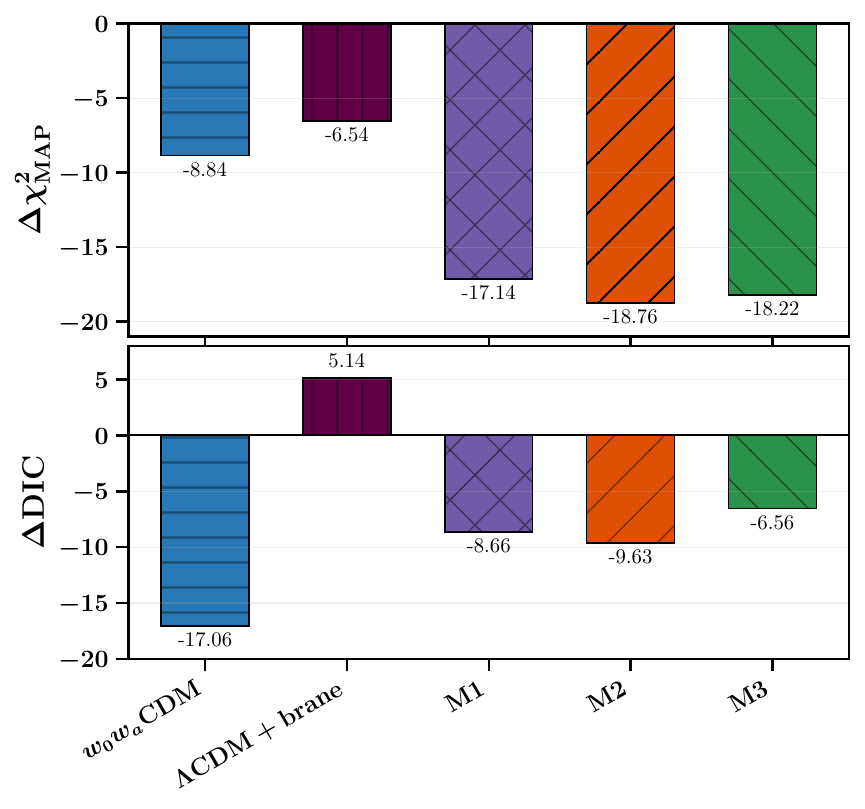}
\caption{$\Delta\chi^2_{\rm MAP}$ and $\Delta{\rm DIC}$ relative to \LCDM. The metastable-brane models provide the largest improvements in the MAP, whereas CPL has the most negative DIC.}
\label{fig:modelcomparison}
\end{figure}

\item The relative fit statistics are summarized in Table~\ref{tab:modelcomparison} and Fig.~\ref{fig:modelcomparison}. All three metastable models improve the MAP likelihood relative to \LCDM, with \begin{equation} 
\Delta\chi^2_{\rm MAP} \ = \ 
\begin{cases}
-17.14 \quad &\text{for \ M1}\, ,\\
-18.76 \quad &\text{for \ M2}\, ,\\
-18.22 \quad &\text{for \ M3}\, .
\end{cases}
\end{equation}
Their corresponding $\Delta{\rm DIC}$ values are $-8.66$, $-9.63$, and $-6.56$, respectively. All three metastable models therefore remain preferred over \LCDM\ under the DIC, which accounts for effective model complexity. M2 gives the best MAP fit among all models considered and the lowest DIC among the metastable models.

The minimal phantom-brane model with a cosmological constant improves the MAP likelihood by $\Delta\chi^2_{\rm MAP}=-6.54$, but its $\Delta{\rm DIC}=+5.14$ indicates that the likelihood gain does not compensate for the additional effective complexity, leaving \LCDM\ preferred under this criterion. By contrast, $w_0w_a$CDM gives $\Delta\chi^2_{\rm MAP}=-8.84$ and $\Delta{\rm DIC}=-17.06$. Although its MAP likelihood improvement is smaller than for M1--M3, $w_0w_a$CDM has the lowest DIC among the models considered. Thus, the metastable-brane models achieve the largest MAP likelihood gains, whereas the DIC favors $w_0w_a$CDM.
\end{enumerate}
Together, the posterior constraints and profile likelihoods support metastable DE decay on the phantom brane. The interplay between decay and gravitational screening produces an effective phantom-divide crossing, providing a physical explanation for the inferred DE evolution while improving the fit to CMB+DESI+DES relative to standard \LCDM.

\section{Concluding Remarks}
\label{sec:summary}

We have investigated metastable dark energy on the minimal phantom brane using CMB observations, DESI DR2 BAO measurements, and the DES-Dovekie Type Ia supernova data. We considered three decay prescriptions: exponential decay of DE (M1), decay of DE into non-baryonic dark matter (M2), and decay of DE into dark radiation (M3). Our model combines two distinct effects: metastable DE decay yields a quintessence-like EoS in the conserved-fluid description, while gravitational screening on the phantom brane leads to an effective phantom behaviour at early times. This interesting combination provides a simple mechanism for crossing the phantom divide at $z \sim 0.4$ without requiring the metastable component itself to be phantom.

For all three metastable prescriptions, the combined data prefer positive decay rates, with $\Gamma/H_0=0.362^{+0.094}_{-0.076}$, $0.539\pm0.125$, and $0.483\pm0.112$ for M1, M2, and M3, respectively, at $68\%$ CL. The profile likelihoods also have minima at positive decay rates and nonzero $\Omega_\ell$, consistent with the marginalized posteriors. The reconstructed metastable-component histories satisfy $w_\phi>-1$, whereas the total effective EoS $w_{\rm DE}$ crosses the phantom divide $w_{\rm DE} = -1$ at $z\simeq0.3$--$0.4$ in the representative histories. Thus braneworld screening drives phantom behaviour at intermediate redshift, while DE decay produces non-phantom behaviour towards the present epoch. The minimal phantom brane with a cosmological constant does not reproduce the same late-time behaviour as our model because DE in the former is simply the $\Lambda$-term which remains constant in time and does not decay.

All three metastable models improve the MAP likelihood. Relative to \LCDM, M1, M2, and M3 give $\Delta\chi^2_{\rm MAP}=-17.14$, $-18.76$, and $-18.22$, respectively. Their corresponding $\Delta{\rm DIC}$ values are $-8.66$, $-9.63$, and $-6.56$, so the preference over \LCDM\ persists under this criterion. M2 gives the largest MAP likelihood improvement among the models considered. The minimal phantom-brane model improves the MAP likelihood but gives $\Delta{\rm DIC}=+5.14$, while CPL gives a smaller MAP improvement  of $\Delta\chi^2_{\rm MAP}=-8.84$ (compared to the metastable models) but the strongest DIC preference, $\Delta{\rm DIC}=-17.06$. Thus, the metastable models achieve larger likelihood gains, whereas CPL is preferred under the DIC.

The models also differ in their impact on structure. M1 and M3 retain $\Omega_{m}$ and $S_8$ close to the \LCDM values, whereas M2 predicts a substantially larger matter fraction and a lower clustering amplitude, with $\Omega_{m}=0.441\pm0.038$ and $S_8=0.733^{+0.016}_{-0.020}$. These shifts accompany the transfer of energy from DE to DM, as DE decay directly changes the dark-matter abundance and growth. All three models yield similar Hubble constants, $H_0\simeq67.6\,\mathrm{km\,s^{-1}\,Mpc^{-1}}$. The sound horizon remains nearly unchanged, and the inferred $H_0$ values lie slightly below the \LCDM\ estimate.

We also constructed an equivalent canonical scalar-field description of the metastable component and compared several potential forms with the reconstructed posterior branches. Among the quadratic, quartic, exponential, and axion forms tested, the exponential form best approximates the reconstructed potentials in all three models. The quadratic ranks second for M1, while the quartic ranks second for M2 and M3. This comparison identifies useful approximations over the reconstructed field intervals; it does not establish their cosmological likelihood ranking or determine the potential beyond those intervals.

Our results provide a physical realization of effective phantom-divide crossing through metastable DE decay and braneworld screening. Extending the analysis to full-shape galaxy clustering data offers a complementary test, particularly of the high matter density and low clustering amplitude preferred by M2. Earlier metastable-DE analyses have demonstrated the sensitivity of these measurements to the DM decay channel. A consistent treatment of both the interaction and braneworld effects on structure growth can therefore test whether the preference for decay persists beyond the present data combination.

\begin{acknowledgments}
PM and AS acknowledge funding from the Korea Astronomy and Space Science Institute (KASI) through the project “Research on the Principles of the Accelerating Expansion of the Universe” (Project Code: 2026183201).  VS thanks the Anusandhan National Research
Foundation (ANRF), India, for the National Science 
Chair Professorship, which provided partial funding for
this work. YS acknowledges funding from the National Academy of Sciences of Ukraine
under Project 0126U000353. The authors acknowledge the use of computational resources of the HPC cluster \textit{Jindeok} at KASI. 
\end{acknowledgments}

\bibliography{references}

\end{document}